  \documentclass[graphicx,amssymb,amsmath,enumerate,11pt]{report}
  \usepackage{fancyhdr}
  \fancyheadoffset[]{0.1cm}
  \usepackage{epsfig}
  \usepackage{url}

\usepackage{color}
\usepackage{multirow}
\usepackage{subfig}
\usepackage{amssymb}

\newcommand\mchapter[2]{\chapter*{#1}
\vskip -0.5cm \noindent {\it \LARGE #2}

\addcontentsline{toc}{chapter}{#1\\{\normalsize\it #2}}}

\def\lsim{\raise0.3ex\hbox{$\;<$\kern-0.75em\raise-1.1ex
\hbox{$\sim\;$}}}
\def\gsim{\raise0.3ex\hbox{$\;>$\kern-0.75em\raise-1.1ex
\hbox{$\sim\;$}}}

\begin{document}

 \rhead{\bfseries Solar Neutrinos}

 \mchapter{Solar Neutrino Observables Sensitive to Matter Effects}
 {Author:\ H. Minakata$^1$ and C. Pe\~na-Garay$^2$}
 \label{ch-XX:instructions}

\vspace{0.1cm}

\begin{center}
$^1$ {\it Department of Physics, Tokyo Metropolitan University, Hachioji, Tokyo 192-0397, Japan} \\ [6pt]
$^2$ {\it Institut de F\'{\i}sica Corpuscular, CSIC-Universitat de Val{\`e}ncia, Apartado de Correos 22085, E-46071 Val{\`e}ncia, Spain}
\end{center}

\vspace{0.3cm}

\begin{center}
{\bf Abstract}
\end{center}
We discuss constraints on the coefficient $A_{MSW}$ which is introduced to simulate the effect of weaker or stronger matter potential for electron neutrinos with the current and future solar neutrino data. The currently available solar neutrino data leads to a bound $A_{MSW} = 1.47^{-0.42}_{+0.54} (^{-0.82}_{+1.88}) $ at 1$\sigma$ (3$\sigma$) CL, which is consistent with the Standard Model prediction $A_{MSW} = 1$. For weaker matter potential ($A_{MSW} < 1$), the constraint which comes from the flat $^8$B neutrino spectrum is already very tight, indicating the evidence for matter effects. Whereas for stronger matter potential ($A_{MSW} > 1$), the bound is milder and is dominated by the day-night asymmetry of $^8$B neutrino flux recently observed by Super-Kamiokande.  
Among the list of observable of ongoing and future solar neutrino experiments, we find that (1) an improved precision of the day-night asymmetry of $^8$B neutrinos, (2) precision measurements of the low energy quasi-monoenergetic neutrinos, and (3)  the detection of the upturn of the $^8$B neutrino spectrum at low energies, are the best choices to improve the bound on $A_{MSW}$. 

\section{Introduction} 
\label{introduction} 

Neutrino propagation in matter is described by the Mikheyev-Smirnov-Wolfenstein 
(MSW) theory \cite{MSW}. It was successfully applied to solve the solar neutrino problem \cite{bahcall}, the discrepancy between the data \cite{Cl,Ga,SK,SNO,Borexino} and the theoretical prediction of solar neutrino flux \cite{SSM}, which blossomed into the solution of the puzzle, the large-mixing-angle (LMA) MSW solution.  The solution is in perfect agreement with the result obtained by KamLAND \cite{KamLAND} detector which measured antineutrinos from nuclear reactors, where the flavor conversion corresponds to vacuum oscillations with sub-percent corrections due to matter effects. 

The MSW theory relies on neutrino interaction with matter dictated by the 
standard electroweak theory and the standard treatment of refraction 
which is well founded in the theory of refraction of light. 
Therefore, it is believed to be on a firm basis. 
On the observational side it predicts a severer reduction of the solar neutrino flux 
at high energies due to the adiabatic flavor transition in matter than at low energies 
where the vacuum oscillation effect dominates. Globally, the behavior is indeed 
seen in the experiments observing $^8$B solar neutrinos at high energies \cite{SK,SNO} and 
in radiochemical experiments detecting low energy $pp$ and 
$^7$Be neutrinos \cite{Cl,Ga}, and more recently by the direct measurement of $^7$Be neutrinos by Borexino \cite{Borexino}. 
For a summary plot of the current status of high and low energy solar neutrinos, see the review of solar neutrinos in this series. 
Therefore, one can say that the MSW theory is successfully confronted with the available experimental data. 

Nevertheless, we believe that further test of the MSW theory is worth pursuing. First of all, it is testing the charged current (CC) contribution to the index of refraction of neutrinos of the Standard Model (SM), which could not be tested anywhere else. 
Furthermore, in analyses of future experiments to determine $\delta$ and the mass hierarchy, the MSW theory is usually assumed to disentangle the genuine effect of CP phase $\delta$ from the matter effect. Therefore, to prove it to the accuracy required by measurement of $\delta$ is highly desirable to make discovery of CP violation robust in such experiments that could have matter effect contamination. This reasoning was spelled out in \cite{mina-venice07}. Since the survival probability $P_{ee}$ does not depend on $\delta$ solar neutrinos provide with us a clean environment for testing the theory of neutrino propagation in matter.

We notice that in solar neutrinos, the transition from low to high energy behaviors mentioned above has not been clearly seen in a single experiment in a solar-model independent manner. The Borexino and KamLAND experiments tried to fill the gap by observing $^8$B neutrinos at relatively low energies \cite{Borexino-8B,KamLAND-8B}. SNO published the results of analyses with lower threshold energy of 3.5 MeV \cite{SNO-lowE,SNO-I-II-III}, and the similar challenge is being undertaken by the Super-Kamiokande (SK) group \cite{SK-DN}. In addition to $^7$Be, a new low-energy neutrino line, $pep$ neutrinos, was observed by Borexino \cite{Borexino-pep}.
Recently, the SK group announced their first detection of the day-night asymmetry of $^8$B neutrinos \cite{SK-DN}. As we will see in section~\ref{constraint} it gives a significant impact on our discussions. With these new experimental inputs, as well as all the aforementioned ones, it is now quite timely to revisit the question of how large deviation from the MSW theory is allowed by data.

In this paper, we perform such a test of the theory of neutrino propagation in the environments of solar and Earth matter. For this purpose, we need to specify the framework of how deviation from the MSW theory is parametrized. We introduce, following \cite{bari}, the parameter $A_{MSW}$ defined as the ratio of the effective coupling of weak interactions measured with coherent neutrino matter interactions in the forward direction to the Fermi coupling constant $G_F$. 
We first analyze the currently available solar neutrino data to obtain the constraints on $A_{MSW}$, and find that the features of the constrains differ depending upon $A_{MSW} < 1$ or $A_{MSW} > 1$. We will discuss interpretations of the obtained constraints including this feature, and provide a simple qualitative model to explain the bound at $A_{MSW} > 1$, more nontrivial one. We then discuss the question of to what extent the constraints on $A_{MSW}$ can be made stringent by various future solar neutrino observables. 

Our framework of testing the theory of neutrino propagation in matter requires comments. It actually involves the three different ingredients: 
(1) non-SM weak interactions in the forward direction parametrized as $A_{MSW} G_F$, (2) refraction theory of neutrino propagation in matter which includes the resonant enhancement of neutrino flavor conversion \cite{MSW}, (3) electron number densities in the Sun and in the Earth.
However, on ground of well founded refraction theory, and because no problem can be arguably raised in the formulation of the MSW mechanism we do not question the validity of (2). 
We also note that the electron number density in the Sun is reliably calculated by the standard solar model (SSM), and the result is cross checked by helioseismology to an accuracy much better than the one discussed here. We can also take the Earth matter density and chemical compositions calculated by the Preliminary Reference Earth Model (PREM) \cite{PREM} as granted.  It is the case because the Earth matter dependent observable, the day-night variation of solar neutrino flux, is insensitive to the precise profile of the Earth matter density. Therefore, we assume that our test primarily examines the aspect (1), namely, whether neutrino matter coupling in the forward direction receive additional contribution beyond those of SM. 

Are non-SM weak interactions parametrized as $A_{MSW} G_F$ general enough? Most probably not because in many models with new non-SM interactions they have flavor structure. Flavor dependent new neutral current interactions have been discussed in the framework of nonstandard interactions (NSI) of neutrinos \cite{NSI-early}, and constraints on effective neutrino matter coupling were obtained with this setting, e.g., in \cite{Davidson,Biggio}. With solar neutrinos see \cite{friedland04} for discussion of NSI. 
If we denote the elements of NSI as $\varepsilon_{\alpha \beta}$ ($\alpha \beta = e, \mu, \tau$), our $A_{MSW}$ may be interpreted as $A_{MSW} = 1 + \varepsilon_{ee}$, assuming that $\varepsilon_{\alpha \beta} \ll \varepsilon_{ee}$ for $\alpha \neq e, \beta \neq e$. To deal with the fully generic case, however, we probably have to enlarge the framework of constraining the NSI parameters by including other neutrino sources, in particular, the accelerator and atmospheric neutrinos. See section \ref{conclusion} for more comments.

\section{Simple analytic treatment of matter effect dependences}
\label{analytic}

In this section, we give a simple analytic description of how various solar 
neutrino observables depend upon the matter effect. 
It should serve for intuitive understanding of the characteristic features 
which we will see in the later sections. The reader will find a physics discussion in the flavor conversion review of this series. 
In the following, we denote the matter densities inside the Sun and in the Earth as 
$\rho_{S}$ and $\rho_{E}$, respectively. 
Solar neutrino survival or appearance probabilities depend on three oscillation parameters: the solar oscillation parameters ($\theta_{12}$, $\Delta m^2_{21} \equiv m^2_{2} - m^2_{1}$), and $\theta_{13}$. Smallness of the recently measured value of $\theta_{13}$ \cite{T2K,MINOS,DC,DB,RENO} and its small error greatly restricts the uncertainty introduced by this parameter on the determination of matter effects.

To quantify possible deviation from the MSW theory, we introduce the parameter $A_{MSW}$ by replacing the Fermi coupling constant $G_F$ by $A_{MSW} G_F$ \cite{bari}. The underlying assumption behind such simplified framework is that the deviation from the Fermi coupling constant is universal over fermions, in particular up and down quarks.

The survival probability in the absence of the Earth matter effect, 
i.e.,  during the day, is well described by \cite{parke,lim,shi-schramm}
\begin{eqnarray}
P_{ee}^{D} =  \cos^4\theta_{13} \left(  \frac{1}{2} + \frac{1}{2} \cdot \cos2\theta_S \cdot \cos2\theta_{12} \right) +  \sin^4\theta_{13}
\label{Pday}  
\end{eqnarray}
Here $\theta_S$ is the mixing angle at the production point inside the Sun:  
\begin{eqnarray}
\cos 2\theta_S \equiv \cos2\theta_m(\rho_S) 
\label{eq}
\end{eqnarray}
where $\theta_m(\rho)$ is the mixing angle in matter of density $\rho_{S}$, 
\begin{eqnarray}
\cos2\theta_S  = { \cos2\theta_{12} - \xi_{S} \over( 1 -2\xi_{S} \cos 2\theta_{12} + \xi_{S}^2 )^{1/2}}.
\label{tSnusun}
\end{eqnarray}
In (\ref{xi-def}), $\xi_{S}$ is defined as the ratio of the neutrino oscillation 
length in vacuum, $l_\nu$, to the refraction length in matter, $l_0$: 
\begin{eqnarray}
\xi_{S} \equiv \frac{l_\nu}{l_0} &=& 
\frac{2 \sqrt{2} A_{MSW} G_F \rho_{S} Y_e  \cos^2\theta_{13}}{m_N} 
\frac{E}{\Delta m^2} 
\nonumber \\ 
&=& 
0.203  \times A_{MSW} \cos^{2} \theta_{13} 
\left(  \frac{E}{ \mbox{ 1 MeV} } \right)
\left(  \frac{ \rho_{S} Y_e }{ \mbox{\rm 100 g\ cm}^{-3}  } \right),
\label{xi-def}
\end{eqnarray}
where 
\begin{eqnarray}
l_\nu\equiv\frac{4\pi E}{\Delta m^2}, 
\hspace{10mm}
l_0\equiv \frac{2\pi m_N}{\sqrt2 A_{MSW} G_F \rho_{S} Y_e \cos^2\theta_{13}}.
\label{length}
\end{eqnarray}
In (\ref{xi-def}) and (\ref{length}), $\rho_{S}$ is the matter density, $Y_e$ is the number of electrons per nucleon, and $m_N$ is the nucleon mass. In the last term we have used the best fit of the global analysis $\Delta m^2_{21}= 7.5 \times 10^{-5}$ eV$^2$. 
The average electron number densities $\rho_{S} Y_e$ at the production point 
of various solar neutrino fluxes are tabulated in Table~\ref{averageNe}. These numbers serve to show the differences in solar densities probed 
by the different sources of neutrinos, but the precise calculations are correctly done by averaging the survival probability with the production point distribution of the corresponding source \cite{holanda,SSM,bahcall}.

We observe that $P_{ee}^{D}$ in (\ref{Pday}) depends on neutrino energy $E$ and $A_{MSW}$ in the particular combination $A_{MSW} E$. The property may have the following implications to constraints on $A_{MSW}$: 
(1) Since shifting $A_{MSW}$ is equivalent to shifting $E$ our analysis which calculate $\chi^2$ as a function of $A_{MSW}$ is inevitably affected by the whole spectrum. 
(2) Nonetheless, we generically expect that the constraint at $A_{MSW} < 1$ ($A_{MSW} > 1$) principally comes from neutrino spectrum at high (low) energies. 
It appears that the apparently contradictory remarks are both true in view of the results in section \ref{constraint}.
 
\begin{table}[!t]
\centering \caption[]{
Average electron density at the neutrino production region and energy of the relevant pp solar neutrinos fluxes.
Last column shows the ratio of the electron neutrino elastic scattering with electrons cross section to the $\mu$ (or $\tau$) neutrino one. For this calculation, we have assumed a measured electron kinetic energy range of [0.05,0.4], [1,1.4], [0,0.8],  and [5,16] MeV for the $pp$, $pep$, $^7$Be and $^8$B respectively.
\protect\label{tab:density}}
\begin{tabular}{lccc}
\noalign{\bigskip} \hline\hline \noalign{\smallskip}
Source&\multicolumn{1}{c}{$\rho_{S} Y_e ({\rm g\ cm}^{-3}) $}& Energy (MeV) & $\frac{\sigma_\mu}{\sigma_e}$\\
\noalign{\smallskip} \hline \noalign{\smallskip}
$pp$&67.9&$\le$0.42&0.284\\
$pep$&73.8&1.44&0.203\\
${\rm ^7Be}$&86.5&0.86&0.221\\
${\rm ^8B}$&92.5& $\le$16&0.155\\
\noalign{\smallskip} \hline\hline \noalign{\smallskip}
 \noalign{\smallskip}
\end{tabular}
\label{averageNe}
\end{table}

\subsection{Energy spectrum}

Solar neutrino observables taken in a single experiment have not shown an energy dependence yet. The neutrino oscillation parameters are such that we can not expect strong energy dependences.
At low neutrino energies, small $\xi_{S}$,
Eq.~(\ref{Pday}) can be approximated by 
\begin{eqnarray}
P_{ee}^{D} = 
\cos^4\theta_{13} \left[  1 - \frac{1}{2}  \sin^2 2\theta_{12}  \left(1 + \cos 2\theta_{12}  \xi_{S} \right) \right] +  \sin^4\theta_{13} 
\label{Pday-lowE}  
\end{eqnarray}
Whereas at high energies, small $\frac{ 1 }{ \xi_{S} }$, 
the oscillation probability (\ref{Pday}) can be approximated, keeping only 
the first energy dependent term as 
\begin{eqnarray}
P_{ee}^{D} =  
\cos^4\theta_{13} \left[  \sin^2 \theta_{12} +  \frac{1}{4}  \sin^2 2\theta_{12} \cos 2\theta_{12} \left( \frac{ 1 }{ \xi_{S} } \right)^2 \right] +  \sin^4\theta_{13}  
\label{Pday-highE} 
\end{eqnarray}
Notice that the correction to the asymptotic behavior is linear in $A_{MSW}$ 
at low energies while it is quadratic in $A_{MSW}^{-1}$ at high energies. 
It may mean that the energy spectrum at low energies could be more advantageous in tightening up the constraint on $A_{MSW}$ provided that these formulas with leading order corrections are valid. 

It is well known that in the LMA MSW mechanism, $^8$B neutrino spectrum must show an upturn from the asymptotic high energy ($E \gg 10$MeV) to lower energies. The behavior is described by the correction term in (\ref{Pday-highE}) but only at a qualitative level. It indicates that the upturn component in the spectrum is a decreasing function of $A_{MSW}$. On the other hand, at low energies populated by $pp$, $^7$Be, and $pep$ neutrinos, the solar neutrino energy spectrum display vacuum averaged oscillations or decoherence, (\ref{Pday-lowE}). The deviation from this asymptotic low energy limit can be described by the correction term in (\ref{Pday-lowE}) again at the (better) qualitative level. The term depend upon $A_{MSW}$ linearly so that the correction term is an increasing function of $A_{MSW}$. Because of the negative sign in the correction term in (\ref{Pday-lowE}), larger values of $A_{MSW}$ lead to smaller absolute values of $P_{ee}$ in both low and high energy regions.\footnote{
The simpler way to reach the same conclusion is to use the property 
$P_{ee}^{D} (E, s A_{MSW}) = P_{ee}^{D} (s E, A_{MSW})$ mentioned earlier. Then, for larger $A_{MSW}$ ($s > 1$) $P_{ee}^{D}$ corresponds to the one at higher energy. Since $P_{ee}^{D}$ is a monotonically decreasing function of $E$, larger the $A_{MSW}$, smaller the $P_{ee}^{D}$.  
}

To see how accurate is the behavior predicted by the above approximate analytic expressions, we have computed numerically (using the PREM profile) the average 
$\langle \left[ \left(1- r_{\mu / e} \right) P_{ee} + r_{\mu / e} \right]  (E_{e, i}) \rangle $ as a function of electron energy. Here, $\langle O \rangle$ means taking average of $P_{ee}$ over neutrino energies with neutrino fluxes times the differential cross sections integrated over the true electron energy with response function. In the above expression, $r_{\mu / e} \equiv \frac{\sigma_\mu}{\sigma_e}$ with $\sigma_e$ and $\sigma_\mu$ being the cross sections of $\nu_{e} e$ and 
$\nu_{\mu} e$ scattering, respectively. 
The computed results confirm qualitatively the behavior discussed above based on our analytic approximations. Thus, the energy spectrum of solar neutrinos at low and high energies can constrain $A_{MSW}$ in this way, as will be shown quantitatively in Sec.~\ref{constraint}.

\subsection{Day-night variation} 
\label{day-night}

The $\nu_{e}$ survival probability at night during which solar 
neutrinos pass through the earth can be written, 
assuming adiabaticity, as \cite{concha-pena-smi} 
\begin{eqnarray}
P_{ee}^{N} = 
P_{ee}^{D} - \cos 2\theta_{S} \cos^2 \theta_{13} \langle f_{reg} \rangle_{\mbox{zenith} }
\label{Pnight}
\end{eqnarray}
where $P_{ee}^{D} $ is the one given in (\ref{Pday}). 
$f_{reg}$ denotes the regeneration effect in the earth, and 
is given as $f_{reg} = P_{2e} - \sin^2\theta_{12}  \cos^2 \theta_{13} $, where 
$P_{2e}$ is the transition probability of second mass eigenstate to $\nu_{e}$. 
Under the constant density approximation in the earth, 
$f_{reg}$ is given by \cite{concha-pena-smi} 
\begin{eqnarray} 
f_{reg} = \xi_{E}  \cos^2 \theta_{13} \sin^2 2\theta_{E} 
\sin^2 \left[ A_{MSW} a_{E} \cos^2 \theta_{13}  (1-2 \xi_{E}^{-1}\cos^2 \theta_{12}+\xi_{E}^{-2})^{\frac{1} {2} } 
\left( \frac{L}{2} \right) \right] 
\label{freg}
\end{eqnarray}
for passage of distance $L$, where we have introduced 
$a_{E} \equiv  \sqrt{2} G_{F} N_{e}^{earth} =  
\frac{ \sqrt{2} G_{F} \rho_{E} Y_{e E} } {m_N} $.
In ($\ref{freg}$), $\theta_{E}$ and $\xi_{E}$ stand for 
the mixing angle and the $\xi$ parameter [see (\ref{xi-def})] 
with matter density $\rho_{E}$ in the earth. 
Within the range of neutrino parameters allowed by the solar neutrino data, 
the oscillatory term averages to $\frac{1}{2}$ in good approximation when 
integrated over zenith angle. Then, the equation simplifies to
\begin{eqnarray}
\langle f_{reg} \rangle_{\mbox{zenith}} = 
\frac{1} {2} \cos^2 \theta_{13}  \xi_{E}  \sin^2 2\theta_{E}.  
\label{freg-ave}
\end{eqnarray}
At $E=7$ MeV, which is a typical energy for $^8$B neutrinos, 
$\xi_{E}=3.98 \times 10^{-2}$ and 
$\sin 2\theta_{E} = 0.940$ for 
the average density 
$\bar{\rho}_{E} = 5.6 \mbox{g/cm}^3$ 
and the electron fraction $Y_{e E} = 0.5$ in the Earth. 
Then, $\langle f_{reg} \rangle_{\mbox{zenith}}$ is given as 
$ \langle f_{reg} \rangle_{\mbox{zenith}} = 1.72 \times 10^{-2}$ 
for $A_{MSW} = 1$ and $\sin^2 2\theta_{13}=0.089$.  
This result is in reasonable agreement with more detailed estimate using the 
PREM profile \cite{PREM} for the Earth matter density.

We now give a simple estimate of the day-night asymmetry $A_{D N}$ assuming 
constant matter density approximation in the earth, and its $A_{MSW}$ dependence. 
Under the approximation of small regeneration effect $f_{reg} \ll 1$, the day-night asymmetry $A_{DN}$ for the CC number of counts N$_{CC}$ measurement 
is approximately given by 
\begin{eqnarray}
A_{DN}^{CC}  &\equiv&  
\frac { N_{CC}^{N} - N_{CC}^{D}} 
{ \frac{1}{2} [ N_{CC}^{N} + N_{CC}^{D} ] } 
\approx 
-  \frac{ 2 \cos 2\theta_{S} }{  1+ \cos 2\theta_{12} \cos 2\theta_{S}  } 
\langle f_{reg} \rangle_{\mbox{zenith}} 
\label{asym-DN-CC}
\end{eqnarray}
where in the right-hand-side we have approximated $A_{DN}^{CC}$ by the 
asymmetry of survival probabilities in day and in night at an appropriate neutrino 
energy, and ignored the terms of order 
$\langle f_{reg} \rangle_{\mbox{zenith}}^2$. 
Notice that the effects of the solar and the earth matter densities are 
contained only in $\cos 2\theta_{S}$ and 
$\langle f_{reg} \rangle_{\mbox{zenith}}$, respectively. 

At $E=7$ MeV,  
$\xi_{S} = 1.31$, 
$\cos 2\theta_{12} = 0.377$, 
$\cos 2\theta_{S} = - 0.710$, 
and hence 
$A_{DN}^{CC} = 3.41  \times 10^{-2} A_{MSW} \cos^4\theta_{13}$, 
about 3\% day-night asymmetry for $A_{MSW} =1$. 
Note that $\cos^4 \theta_{13}=0.95$ for $\sin^2 2\theta_{13} = 0.1$, so that the 
impact of $\theta_{13}$ on $A_{DN}^{CC}$ give only a minor modification.  
Though based on crude approximations, the value of $A_{DN}^{CC}$ at $A_{MSW}=1$ obtained above is in excellent agreement with the one evaluated numerically for SNO CC measurement.

SNO and SK observes the day-night asymmetry by measurement of CC reactions and elastic scattering (CC+NC), respectively. 
We have computed $A_{DN}$ as a function of $A_{MSW}$ numerically (with PREM profile) without using analytic approximation. The result of $A_{DN}$ scales linearly with $A_{MSW}$ in a good approximation, $A_{DN}^{CC} \approx 0.044 A_{MSW}$.  
Similarly, the day-night asymmetry for elastic scattering measurement can be easily computed. Its relationship to the $A_{DN}^{CC}$ can be estimated in 
the similar manner as in (\ref{asym-DN-CC}), 
\begin{eqnarray}
A_{DN}^{ES}  &\equiv&  
\frac { N_{ES}^{N} - N_{ES}^{D}} 
{ \frac{1}{2} [ N_{ES}^{N} + N_{ES}^{D} ] } 
\approx 
A_{DN}^{CC} \times 
\left[
1 + \frac{ 2 r_{\mu / e} }{ \left( 1- r_{\mu / e} \right)  [P_{ee}^{N} + P_{ee}^{D}]  }
\right]^{-1}, 
\label{asym-DN-ES}
\end{eqnarray}
taking into account the modification due to NC scattering. 
Using approximate values, 
$r_{\mu / e}=\frac{1}{6}$ and $\frac{1}{2} [P_{ee}^{N} + P_{ee}^{D}]  =\frac{1}{3}$
the factor in the square bracket can be estimated to be $\frac{5}{8}$, giving a reasonable approximation for the ratio of $A_{DN}^{ES}$ to $A_{DN}^{CC}$. A better approximation to the computed results of the $A_{MSW}$ dependence of the asymmetry is given by $A_{DN}^{ES} =  0.02 A_{MSW}$. 

\section{Constraints on $A_{MSW}$ by Solar Neutrino Observables} 
\label{constraint}

In this section we investigate quantitatively to what extent $A_{MSW}$ can be 
constrained by the current and the future solar neutrino data. 
The results of our calculations are presented in Fig.~\ref{Deltachi2}, 
supplemented with the relevant numbers in Table~\ref{allowed-region}. 
We will discuss the results and their implications to some details in a step-by-step manner.
We first discuss the constraints by the data currently available (Sec.~\ref{current}). 
Then, we address the question of how the constraint on $A_{MSW}$ can be tightened with the future solar neutrino data, the spectral upturn of $^8$B neutrinos (Sec.~\ref{high-E}), the low energy $^7$Be and $pep$ neutrinos (Sec.~\ref{low-E}), and finally the day-night asymmetry of the solar neutrino flux (Sec.~\ref{day-night-asym}).
We pay special attention to the question of how the constraints on 
$A_{MSW}$ depend upon the significance of these measurements. 

\subsection{Current constraint on $A_{MSW}$} 
\label{current}

We include in our global analyses the KamLAND and all the available solar neutrino data~\cite{Cl,Ga,SK,SNO,Borexino,KamLAND,Borexino-8B,KamLAND-8B,SNO-lowE,SNO-I-II-III,SK-DN}. 
To obtain all the results quoted in this paper we marginalize over the mixing angles $\theta_{12}$ and $\theta_{13}$, the small mass squared difference $\Delta m^2_{21}$, and the solar neutrino fluxes $f_{i}$ \cite{SSM,cpg} imposing the luminosity contraint \cite{luminosity}.
We include in the analysis the  $\theta_{13}$ dependence derived from the 
analysis of the atmospheric, accelerator, and reactor data included in Ref.~\cite{maltoni} as well as the recent measurement of $\theta_{13}$ by \cite{T2K,MINOS,DC,DB,RENO}. The $\chi^2$ used is defined by
\begin{eqnarray}
\chi^2_{\rm global} (A_{MSW})&=& Marg [\chi^2_{\rm solar}(\Delta m^2_{21},\theta_{12},\theta_{13}, A_{MSW}, f_{\rm B}, f_{\rm Be},
f_{pp}, f_{\rm CNO}) \nonumber\\
&+& \chi^2_{\rm KamLAND}(\Delta m^2_{21},\theta_{12},\theta_{13})
~+~ \chi^2_{\rm REACTOR+ATM+ACC}(\theta_{13}) ]\,,
\label{eq:chisquared}
\end{eqnarray}
where $Marg$ implies to marginalize over the parameters shown but not over 
$A_{MSW}$. 
Further details of the analysis methods can be found in Ref.~\cite{cpg}.

 \begin{figure}[bhtp]
  \centerline{
\mbox{\includegraphics[width=4.40in]{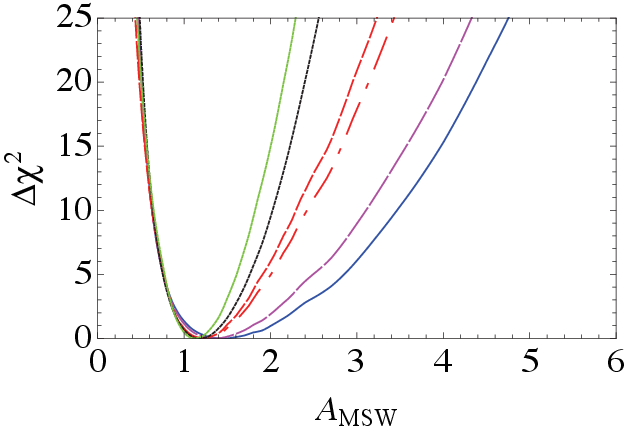} }
  }
\caption{$\Delta \chi^2$ as a function of $A_{MSW}$ for the currently available solar neutrino data (shown in blue solid line) and the various solar neutrino observables expected in the near future (by color lines specified below). The current data include the one from SNO lower energy threshold analysis and SK I-IV. 
In addition to the current constraints on $A_{MSW}$, we show the improved constraints when future solar neutrino data are added one by one: 
3$\sigma$ detection of the SK spectral upturn (magenta dashed line), low energy solar neutrino flux measurements of $^7$Be at 5\% and $pep$ at 3\% (red dash-dotted line), 3$\sigma$ detection of the SK day-night asymmetry (black dotted line).
The red dashed line shows the improved constraints by adding future spectral information at high and low energies. Finally, the global analysis by adding all the spectral 
information data and the day-night data produces the solid green line.
}
 \label{Deltachi2}
  \end{figure}

\begin{table}
\caption[aaa]{
The $\Delta \chi^2$ minimum of $A_{MSW}$, 
the allowed regions of $A_{MSW}$ at 1$\sigma$, and 3$\sigma$ CL are shown 
in the first, second, and third columns, respectively, for the analyses with 
the currently available data (first row), 
the one with spectrum upturn of $^8$B neutrinos at 3$\sigma$ added to the current data (second row), 
the one with $^7$Be and $pep$ neutrinos with 5\% and 3\% accuracies, respectively, added to the current data (third row), 
the one with the new spectral information in the second and the third row  
added to the current data  (fourth row),  and the one with day-night asymmetry of $^8$B neutrinos at 3$\sigma$ added to the current data (fifth row).
The last row presents results of global analysis 
with all the above data.
The numbers in parentheses imply the ones obtained with improved knowledge of 
$\theta_{12}$, see text for details. 
}
\vglue 0.5cm
\begin{tabular}{c|ccc}
\hline
\hline
\ \ Analysis \  &\  $\Delta \chi^2$ minimum  \
            &\ \ allowed region (1$\sigma$) \
            &\ \ allowed region (3$\sigma$)  \\
            \hline
present data   \  &\  $A_{MSW}= 1.47$ \
            &\ 1.05$-$2.01 (1.05$-$2.00) \
            &\ 0.65$-$3.35 (0.65$-$3.27) \\
            \hline
+upturn (3$\sigma$)
                &\ \ 1.34 \
                &\  1.02$-$1.79 (1.02$-$1.76)
                &\  0.65$-$3.00 (0.66$-$2.88)  \\
\hline
+$^7$Be (5\%), $pep$ (3\%)
                &\ \ 1.25
                &\  0.97$-$1.53 (0.97$-$1.52)
                &\  0.65$-$2.34 (0.65$-$2.31) \\
\hline
+spectral shape
                &\ \ 1.22
                &\  0.97$-$1.49 (0.97$-$1.46)
                &\  0.65$-$2.23 (0.65$-$2.12) \\
\hline
+A$_{DN}$ (3$\sigma$)
                &\ \ 1.17
                &\  0.96$-$1.43 (0.96$-$1.42)
                &\  0.66$-$1.98 (0.66$-$1.97) \\
\hline
+all
                &\ \ 1.12  
                &\ 0.95$-$1.33 (0.95$-$1.32) 
                &\ 0.67$-$1.78 (0.67$-$1.73) \\
\hline
\hline
\end{tabular}
\label{allowed-region}
\vglue 0.5cm
\end{table}


The currently available neutrino data (blue solid line), which include SNO lower energy threshold data \cite{SNO-lowE,SNO-I-II-III} and SK IV \cite{SK-DN}, do not allow a very precise determination of the $A_{MSW}$ parameter. A distinctive feature of the $\Delta \chi^2$ parabola shown in Fig.~\ref{Deltachi2} is the asymmetry between the small and large $A_{MSW}$ regions. At $A_{MSW} < 1$ the parabola is already fairly steep, and the ``wall'' is so stiff that can barely be changed by including the future data.  
While at $A_{MSW} > 1$ the slope is relatively gentle. More quantitatively, 
$A_{MSW} = 1.47^{-0.42}_{+0.54} (^{-0.82}_{+1.88}) $ at 1$\sigma$ (3$\sigma$) CL. 
The best fit point with the present data is significantly larger than unity, $A_{MSW} =1.47$. It was 1.32 before and have driven to the larger value mostly by the new SK data which indicates a stronger matter effect than those expected by the MSW LMA region preferred by the KamLAND data. The larger best fit value could also partly be due to an artifact of the weakness of the constraint in $A_{MSW} > 1$ region. Notice that the Standard Model MSW theory value $A_{MSW}=1$ is off from the 1$\sigma$ region but only by a tiny amount, as seen in Table~\ref{allowed-region}. Let us understand these characteristics. 

The lower bound on $A_{MSW}$ mostly comes from the SK and the SNO data which shows that $^8$B neutrino spectrum at high energies is well described by the adiabatic LMA MSW solution ($A_{MSW}=1$). The energy spectrum is very close to a flat one with $P_{ee}$  which can be approximated by $\sin^2 \theta$ with corrections due to the contribution of the energy dependent term (see (\ref{Pday-highE})). The value is inconsistent with the vacuum oscillation, and hence the point $A_{MSW}=0$ is highly disfavored, showing the evidence for the matter effect.  

One would think that the upper bound on $A_{MSW}$ should come from either the low energy solar neutrino data, or the deviation from the flat spectra at high energies. But, we still lack precise informations on low energy solar neutrinos, and the spectral upturn of $^8$B neutrinos has not been observed beyond the level in \cite{Borexino-8B,KamLAND-8B}. Then, what is the origin of the upper bound $A_{MSW} < 2$ at about $1 \sigma$ CL?
We argue that it mainly comes from the day-night asymmetry of $^8$B neutrino flux which is contained in the binned data of SK and SNO. Recently, the SK collaboration reported a positive indication of the day-night asymmetry though the data is still consistent with no asymmetry at 2.3$\sigma$ CL \cite{SK-DN}. 

To show the point, we construct a very simple model for $\Delta \chi^2$ for the day-night asymmetry $A_{DN}^{ES}$. It is made possible by the approximate linearity of $A_{DN}^{ES}$ to $A_{MSW}$. Let us start from the data of day-night asymmetry at SK I-IV obtained with the D/N amplitude method \cite{SK-DN}: 
$A_{DN}^{ES} = (2.8 \pm 1.1 \pm 0.5)$\%, giving the total error 1.2\% if added in quadrature. The expectation of $A_{DN}^{ES}$ by the LMA solution is $A_{DN}^{ES} = A_{MSW} \times 2.1$\% for $\Delta m^2_{21} = 7.6 \times 10^{-5}$ eV$^2$. Then, one can create an approximate model $\Delta \chi^2$ as 
$\Delta \chi^2 = [ \left( A_{DN} - 2.8\% \right) / 1.2\%]^2 = 3.1 \left( A_{MSW} - 1.3 \right)^2$. 

Despite admittedly crude nature it seems to capture the the qualitative features of $\Delta \chi^2$ with the current data (blue solid line) in Fig.~\ref{Deltachi2} in region $A_{MSW} > 1$. It is true that it predicts a little too steep rise of $\Delta \chi^2$ and leads to $\Delta \chi^2 \simeq 22$ at $A_{MSW}=4$, whereas $\Delta \chi^2 \simeq 15$ in Fig.~\ref{Deltachi2}. In the actual numerical analysis for Fig.~\ref{Deltachi2}, however, $\Delta \chi^2$ parabola can naturally become less steep because various other parameters are varied to accommodate such a large values of $A_{MSW}$. Therefore, we find that about 2$\sigma$ evidence of $A_{DN}^{ES}$  in the SK data is most likely the main cause of the sensitivity to $A_{MSW}$ in the region $A_{MSW} > 1$.  
The simple model cannot explain the behavior of $\Delta \chi^2$ in region $A_{MSW} < 1$ in Fig.~\ref{Deltachi2}, because the other more powerful mechanism is at work to lead to stronger bound on $A_{MSW}$, as discussed above. 

To what extent an improved knowledge of $\theta_{12}$ affects $A_{MSW}$? It was suggested that a dedicated reactor neutrino experiment can measure $\sin^2 \theta_{12}$ to $\simeq$2\% accuracy \cite{MNTZ05,bandyopadhyay05}. It is also expected that precision measurement of $pp$ spectrum could improve the accuracy of $\theta_{12}$ determination to a similar extent \cite{cpg}. Therefore, it is interesting to examine to what extent an improved knowledge of $\theta_{12}$ affects the constraint on $A_{MSW}$. Therefore, we re-compute the $\Delta \chi^2$ curves presented in Fig.~\ref{Deltachi2} by adding the artificial term $(\sin^2 \theta_{12} - BEST)^2 / 0.02$ in the $\Delta \chi^2$ assuming 2\% accuracy in $\sin^2 \theta_{12}$ determination. The result of this computation is given in Table~\ref{allowed-region} in parentheses. As we see, size of the effect of improved $\theta_{12}$ knowledge is not very significant. 

\subsection{Spectrum of solar neutrinos at high energies} 
\label{high-E}


Evidence for the upturn of $^8$B neutrino spectrum must contribute to constrain the larger values of $A_{MSW}$ because $A_{MSW}$ could be very large without upturn, if day-night asymmetry is ignored. We discuss the impact on $A_{MSW}$ of seeing the upturn in recoil electron energy spectrum with 3 $\sigma$ significance, which we assume to be in the region $E_{e} \geq 3.5$ MeV. To calculate $\Delta \chi^2$ we assume the errors estimated by the SK collaboration \cite{SK-DN}. 
Adding the simulated data to the currently available data set produces the magenta dashed line in Fig.~\ref{Deltachi2}. 
We find a 25\% reduction of the 3$\sigma$ allowed range, 
$A_{MSW} = 1.34^{-0.32}_{+0.45} (^{-0.69}_{+1.66})$
at 1$\sigma$ (3$\sigma$) CL. 
We can see that it does improve the upper bound on $A_{MSW}$, for which the current constraint (blue solid line) is rather weak, but the improvement in the precision of $A_{MSW}$ is still moderate.

Some remarks are in order about the minimum point of $\Delta \chi^2$. 
The best fit point with the present data is at $A_{MSW} > 1$ as we saw above.
For the analysis with future data discussed in this and the subsequent subsections, we assume that the $\Delta \chi^2$ minimum is always at $A_{MSW}=1$ for simulated data. 
Therefore, the analysis with the present plus simulated data tends to pull the 
$\Delta \chi^2$ minimum toward smaller values of $A_{MSW}$, and at the same time 
make the $\Delta \chi^2$ parabola narrower around the minimum.  
By conspiracy between these two features the current constraint (blue solid line) 
is almost degenerate to the other lines at $A_{MSW} < 1$, the ones with spectral 
upturn (magenta dashed line) and low energy neutrinos (red dash-dotted line).
These features can be observed in Fig.~\ref{Deltachi2} and in Table~\ref{allowed-region}.

\subsection{Spectrum of solar neutrinos at low energies} 
\label{low-E}

Now, let us turn to the low energy solar neutrinos, $^7$Be and $pep$ lines. The Borexino collaboration have already measured the $^7$Be neutrino-electron scattering rate to an accuracy of $\simeq \pm 5$\% \cite{Borexino}, which we assume throughout this section. 
For $pep$ neutrinos we assume measurement with 3\% precision in the future.  
See \cite{Borexino-pep} for the first observation of $pep$ neutrinos, and its current status of the uncertainties. 

The measurement of the $pep$ flux has two important advantages, when compared to the $^7$Be flux, in determining $A_{MSW}$: 
a) the neutrino energy is higher, 1.44 MeV, so the importance of the solar matter effects is larger, 
b) the uncertainty in the theoretical estimate is much smaller. Firstly, the ratio of the $pep$ to the $pp$ neutrino flux is robustly determined by the SSM calculations, so it can be determined more accurately than the individual fluxes because the ratio depends only weakly on the solar astrophysical inputs. Secondly, a very precise measurement of the $^7$Be flux, with all the other solar data and assuming energy conservation (luminosity constraint), leads to a very precise determination of the $pp$ and $pep$ flux, at the level of $\sim$ 1\% accuracy \cite{cpg}. 
%
%
On the other hand, to determine $^7$Be flux experimentally, we have to use the SSM flux to determine the neutrino survival probability, and therefore, the uncertainties in the theoretical estimate \cite{SSM} limit the precision of the $^7$Be flux measurement.


The red dash-dotted line in Fig.~\ref{Deltachi2} shows the result of the combined analysis of future low energy data, an improved $^7$Be measurement with 5\% precision and a future $pep$ measurement with 3\% precision, added to the current data. 
The obtained constraint on $A_{MSW}$ is:
$A_{MSW} = 1.25 \pm 0.28  (^{-0.60}_{+1.09}) $ 
at 1$\sigma$ (3$\sigma$) CL. 
The resultant constraint on $A_{MSW}$ from above is much more powerful than the one obtained with spectrum upturn of high energy $^8$B neutrinos at 3$\sigma$. 

By having solar neutrino spectrum informations both at high and low energies it is tempting to ask how tight the constraint become if we combine them. The result of this exercise is plotted by the red dashed line in Fig.~\ref{Deltachi2} and is also given in Table~\ref{allowed-region}. The resultant constraint on $A_{MSW}$ is: 
$A_{MSW} = 1.22^{-0.25}_{+0.27}  (^{-0.57}_{+1.01}) $ 
at 1$\sigma$ (3$\sigma$) CL.

\subsection{Day-night asymmetry} 
\label{day-night-asym}

To have a feeling on to what extent constraint on $A_{MSW}$ can be tightened by possible future measurement, we extend the simple-minded model discussed in Sec.~\ref{current}, but with further simplification of assuming $A_{MSW} = 1$ as the best fit. Let us assume that the day-night asymmetry $A_{DN}^{ES}$ can be determined with $(2/N)$\% accuracy, an evidence for the day-night asymmetry at $N \sigma$ CL. Then, the appropriate model $\Delta \chi^2$ is given under the same approximations as in Sec.~\ref{current} as 
$\Delta \chi^2  = N^2 \left( A_{MSW} - 1 \right)^2$. 
We boldly assume that the day-night asymmetry at $3 \sigma$ CL would be a practical goal in SK. It predicts $\Delta \chi^2 = 9 \left( A_{MSW} - 1 \right)^2$, 
which means that $A_{MSW}$ can be constrained to the accuracy of 33\% uncertainty at 1$\sigma$ CL.

Now, we give the result based on the real simulation of data. The black dotted line in Fig.~\ref{Deltachi2} show the constraint on $A_{MSW}$ obtained by future $3 \sigma$ CL measurement of the day-night asymmetry, which is added to the present solar neutrino data. As we see, the day-night asymmetry is very sensitive to the matter potential despite 
our modest assumption of $3 \sigma$ CL measurement of $A_{DN}^{ES}$.\footnote{
Given the powerfulness of the day-night asymmetry for constraining $A_{MSW}$, 
it is highly desirable to measure it at higher CL in the future. 
Of course, it would be a challenging task, and probably requires a megaton class water Cherenkov or large volume liquid scintillator detectors with solar neutrino detection capability. They include, for example, Hyper-Kamiokande \cite{Hyper-K},  UNO \cite{UNO}, or the ones described in \cite{LAGUNA}.
}
%
The obtained constraint on $A_{MSW}$ is: 
$A_{MSW} = 1.17^{-0.21}_{+0.26} (^{-0.51}_{+0.81})$ 
at 1$\sigma$ (3$\sigma$) CL.
The obtained upper bound on $A_{MSW}$ is actually stronger than the one expected by our simple-minded model $\Delta \chi^2$. Apart from the shift of the bast fit to a larger value of $A_{MSW}$, the behavior of $\Delta \chi^2$ is more like $\Delta \chi^2 \approx 14 \left( A_{MSW} - 1 \right)^2$ in the region $A_{MSW} > 1$.
It can also been seen in Fig.~\ref{Deltachi2} that the upper bound on $A_{MSW}$ due to the day-night asymmetry at 3$\sigma$ CL (black dotted line) is stronger than the one from combined analysis of all the expected measurements of the shape of the spectrum (red dashed line) discussed at the end of Sec.~\ref{low-E}.

\subsection{Global analysis} 
\label{global}

We now discuss to what extent the constraint on $A_{MSW}$ can become stringent 
when all the data of various observable are combined. 
The solid green line in Fig.~\ref{Deltachi2} shows the constraint on $A_{MSW}$ 
obtained by the global analysis combining all the data sets 
considered in our analysis. The obtained sensitivity reads 
$A_{MSW} = 1.12^{-0.17}_{+0.21} (^{-0.45}_{+0.66})$
at 1$\sigma$ (3$\sigma$) CL. 
Therefore, the present and the future solar neutrino data, under the assumptions of 
the accuracies of measurement stated before, can constrain $A_{MSW}$ 
to $\simeq$15\% (40\%) at 1$\sigma$ (3$\sigma$) CL from below, and 
to $\simeq$20\% (60\%) at 1$\sigma$ (3$\sigma$) CL from above. 
If we compare this to the current constraint 
$A_{MSW} = 1.47^{-0.42}_{+0.54} (^{-0.82}_{+1.88}) $ the improvement of the errors for $A_{MSW}$ over the current precision is, very roughly speaking, a factor of $\simeq 1.5 - 2$ in region $A_{MSW} < 1$, and it is a factor of $\simeq 2$ at $A_{MSW} > 1$. 
Noticing that the efficiency of adding more data to have tighter constraint at $A_{MSW} < 1$ is weakened by shift of the minimum of $\Delta \chi^2$, improvement of the constraint on $A_{MSW}$ is more significant at $A_{MSW} > 1$.

\section{Summary} 
\label{conclusion}

In this paper, we have discussed the question of to what extent tests of the MSW theory can be made stringent by various solar neutrino observables. First, we have updated the constraint on $A_{MSW}$, the ratio of the effective coupling constant of neutrinos to $G_F$, the Fermi coupling constant with the new data including SNO $^8$B spectrum and SK day-night asymmetry. Then, we have discussed in detail how and to what extent the solar neutrino observable in the future tighten the constraint on $A_{MSW}$. 

The features of the obtained constraints can be summarized as follows:

\begin{itemize}

\item 
Interpretation of solar neutrino data at high energies by the vacuum oscillation is severely excluded by the SNO and SK experiments, which leads to a strong and robust lower bound on $A_{MSW}$. 
On the other hand, the day-night asymmetry at $\simeq 2\sigma$ level observed by SK dominates the bound at high $A_{MSW}$ side. We find that present data lead to 
$A_{MSW} = 1.47^{-0.42}_{+0.54} (^{-0.82}_{+1.88}) $ 
at 1$\sigma$ (3$\sigma$) CL. The Standard Model prediction $A_{MSW} = 1$ is outside the 1$\sigma$ CL range but only by tiny amount.

\item 
We have explored the improvements that could be achieved by solar neutrinos experiments, ongoing and in construction.  We discussed three observables that are sensitive enough to significantly improve the limits on  $A_{MSW}$, particularly in the region $A_{MSW} > 1$: 
a) upturn of the $^8$B solar neutrino spectra at low energies at 3$\sigma$ CL, 
b) high precision measurement of mono-energetic low energy solar neutrinos, $^7$Be (5\% precision) and $pep$ (3\% precision) neutrinos, and 
c) day-night asymmetry of the  $^8$B solar neutrino flux at 3$\sigma$ CL. 
They lead to the improvement of the bound as follows: \\
\noindent
a) $A_{MSW} = 1.34^{-0.32}_{+0.45} (^{-0.69}_{+1.66})$ at 1$\sigma$ (3$\sigma$) CL. \\
\noindent
b) $A_{MSW} = 1.25 \pm 0.28  (^{-0.60}_{+1.09})$ at 1$\sigma$ (3$\sigma$) CL.\\
\noindent
c) $A_{MSW} = 1.17^{-0.21}_{+0.26} (^{-0.51}_{+0.81})$ at 1$\sigma$ (3$\sigma$) CL.\\
\noindent 
It could be expected that future measurement by SNO$^+$ \cite{SNO+} and KamLAND \cite{KamLAND-lowE} may detect spectrum modulation of B neutrinos at low energies at CL higher than $3\sigma$. 

Finally, by combining all the data set we have considered we obtain 
$A_{MSW} = 1.12^{-0.17}_{+0.21} (^{-0.45}_{+0.66})$ 
at 1$\sigma$ (3$\sigma$) CL. 

\item 
As mentioned in section \ref{introduction}, the issue of effective neutrino matter coupling in a wider context may be better treated in the framework of NSI. If we think about the extended setting together with accelerator and atmospheric neutrino measurement to look for effects of NSI, the off-diagonal elements $\varepsilon_{\alpha \beta}$ ($\alpha \neq \beta$) can be better constrained by long-baseline experiments. In fact, in a perturbative treatment with small parameter $\epsilon \equiv \frac{\Delta m^2_{21}}{\Delta m^2_{31}}$ with the assumption $\varepsilon_{\alpha \beta} \sim \epsilon$, the terms with $\varepsilon_{e \mu}$ and $\varepsilon_{e \tau}$ are of second order in $\epsilon$, while $\varepsilon_{ee}$ comes in only at third order in $\epsilon$ \cite{NSI-perturbation}. The analyses show that the sensitivity to $\varepsilon_{ee}$ is indeed lower at least by an order of magnitude compared to the ones to $\varepsilon_{e \mu}$ or $\varepsilon_{e \tau}$. See the analysis in \cite{NSI-Madrid}, and the references cited therein. It is also known that $\varepsilon_{\mu \tau}$ can be severely constrained by atmospheric neutrinos \cite{NSI-atm}. Hence, we feel that the solar neutrinos are still a powerful and complementary probe for $\varepsilon_{ee}$ in such the extended setting.

\end{itemize}

In conclusion, testing the theory of neutrino propagation in matter deserves further endeavor. The lack of an accurate measurement of the matter potential felt by solar neutrinos reflects the fact that solar neutrino data only do not precisely determine the mass square splitting. The good match of the independently determined mass square splitting by solar neutrino data and by reactor antineutrino data will confirm the Standard Model prediction of the relative index of refraction of electron neutrinos to the other flavor neutrinos. The lack of match of both measurements would point to new physics like the one tested here.

\vspace{0.5cm}

We thank the Galileo Galilei Institute for Theoretical Physics and 
the organizers of the workshop ``What is $\nu$?'' for warm hospitality.
C.P-G is supported in part by the Spanish MICINN grants FPA-2007-60323, FPA2011-29678, the Generalitat Valenciana grant PROMETEO/2009/116 and the ITN INVISIBLES (Marie Curie Actions, PITN-GA-2011-289442).
H.M. is supported in part by KAKENHI, Grant-in-Aid for Scientific Research No. 23540315, Japan Society for the Promotion of Science.

\end{document}